\documentclass[floatfix,aps,tightenlines,prd,nofootinbib,notitlepage,superscriptaddress,12pt]{revtex4-1}

\usepackage{amsmath}
\usepackage{amssymb}
\usepackage{color}
\usepackage{graphicx}

\newcommand{\Chris}[3]{\ensuremath{\Gamma^{#1}_{\cdot \, #2 #3}}}
\newcommand{\Christilde}[3]{\ensuremath{ \tilde{\Gamma}^{#1}_{\cdot \, #2 #3}}}

\newcommand{\J}{\ensuremath{\mathrm{J}}}

\newcommand{\h}{\ensuremath{\mathrm{H}}}

\newcommand{\g}{\ensuremath{\mathrm{g}}}
\newcommand{\R}{\ensuremath{\mathcal{R}}}

\newcommand{\x}{\tilde{r}}

\begin{document}
\title{Horizonless, singularity-free, compact shells  satisfying NEC }
\author{Karthik H. Shankar}
\affiliation{Initiative for Physics and Mathematics of Neural Systems, \\ Center for memory and Brain. \\ Boston University.}

\begin{abstract}

Gravitational collapse singularities are undesirable, yet inevitable to a large extent in General Relativity. When matter satisfying null energy condition (NEC) collapses to the extent a closed trapped surface is formed, a singularity is inevitable according to Penrose's singularity theorem.  Since positive mass vacuum solutions are generally black holes with trapped surfaces inside the event horizon, matter cannot collapse to an arbitrarily small size without generating a singularity. However, in modified theories of gravity where positive mass vacuum solutions are naked singularities with no trapped surfaces, it is reasonable to expect that matter can collapse to an arbitrarily small size without generating a singularity. Here we examine this possibility in the context of a modified theory of gravity with torsion in an extra dimension. We study singularity-free static shell solutions to evaluate the validity of NEC on the shell. We find that with sufficiently high pressure, matter can be collapsed to arbitrarily small size without violating NEC and without producing a singularity. 

\end{abstract}

\maketitle{}

 \section{Introduction}

How small can a massive object gravitationally collapse without producing a singularity, while still satisfying the Null Energy Condition (NEC)? In the context of General Relativity (GR) the answer is quite clear. In the spherically symmetric case, the vacuum outside the matter distribution is described by the Schwarzschild metric; and once the matter collapses to a size smaller than the Schwarzschild radius, trapped surfaces will form forcing even light rays to fall towards the center creating a black hole. 
Unlike the Schwarzschild black hole, electrically  charged (Reisner-Nordstorm, RN) black holes have a time-like singularity and the region surrounding the singularity is free of trapped surfaces, although there exists closed trapped surfaces farther away from the singularity. So, one could naively expect that an  arbitrarily compact static matter distribution can give rise to a charged black hole-like external geometry. However, according to Penrose's singularity theorem, if the matter obeys null energy condition (NEC), then the mere existence of a closed trapped surface would lead to a singularity \cite{penrose1965gravitational, senovilla20141965}. Hence a RN exterior cannot be produced by an arbitrarily compact singularity-free matter distribution without violating NEC. 

Although cosmic censorship hypothesis proposes,  for philosophical reasons, that any singularity should be shielded by trapped surfaces and event horizon \cite{penrose1999question}, it is in principle possible to produce  naked singularities in gravitational collapse \cite{shapiro1991formation,wald1999gravitational} in GR. Exact solutions with naked singularities can be realized in GR by simply introducing massless scalar charge, given by the JNW metric \cite{janis1968reality}. There are distinct observational differences in gravitational lensing phenomena that can distinguish such solutions from the black hole solutions \cite{virbhadra2002gravitational}; hence theoretically exploring such solutions is very important. 

Any theory of gravity that leads to positive mass naked singularity solutions can be expected to accommodate arbitrarily compact singularity-free matter distribution satisfying NEC. In this paper, we focus on one such theory  that minimally modifies GR by introducing torsion in an extra dimension so as to hide the extra dimension from the visible dimensions \cite{shankar2012metric}, abbreviated in this paper as \emph{THED} --\emph{T}orsion to \emph{H}ide \emph{E}xtra \emph{D}imension.  Here we restrict our focus on the mere existence of singularity-free solutions; so we do not study a detailed  gravitational collapse, rather we focus on static shells with surface density and pressure satisfying the NEC.  Although NEC is the weakest of the energy conditions,  it appears to be the only energy condition in agreement with known macroscopic matter and energy sources in the universe. Any stronger energy condition seems to be violated at high energies or due to quantum effects.  For a comprehensive review of the energy conditions, see \cite{curiel2014primer}.  
Here we shall take the reasonable stance that if  NEC satisfying matter can be compressed to arbitrarily small size without creating trapped surfaces, then collapse singularities can potentially be completely avoided in that theory gravity.

The standard view these days is that quantum gravity effects will prevent the formation of singularity inside black holes. From a string theory perspective, it has been proposed that stable states of intersecting brane configurations can give rise to  singularity-free quantum-fuzzball solutions \cite{mathur2005fuzzball}. Horizonless solutions that appear as black holes at large distances can be constructed from a variety of brane configurations \cite{balasubramanian2008four, rama2013m, rama2011static, rama2014massive}. However, here we shall just consider gravity at the classical level and examine the possibility of arbitrarily compact horizonless singularity-free solutions.

In section 2, we start with a brief review of THED gravity explaining how the field equations are modified from the GR field equations. In section 3, we analyze the properties of positive mass naked singularity solutions obtained from THED. In section 4, we apply Israel's junction conditions \cite{israel1966singular} to derive the surface density and pressure on a static shells. In section 5, we numerically evaluate the validity of NEC on compact shells, and compare the solution from THED gravity to that of RN and JNW geometries. Finally, in section 6, we discuss some implications of THED gravity in generally avoiding gravitational singularities. 

\section{Reviewing $THED$ gravity}

Torsion has been the natural mathematical ingredient needed to formalize the gravitational effects of matter with spin \cite{hehl1976general}, and it plays an essential role in deriving Palatini theories of gravity \cite{olmo2013importance}. It has been argued that the effects of torsion could be strong enough to even avoid big bang singularity in the early universe leading to a singularity-free bounce cosmology \cite{poplawski2012nonsingular} and bubble universes within black holes \cite{poplawski2014universe}. In contrast to other theories with torsion, in THED gravity torsion plays a purely algebraic role. It is not dynamically independent and it does not couple to the spin of matter, rather it is completely determined in terms of the metric to ensure that the extra dimension remains hidden. In this section, we briefly summarize the formulation of this theory, but see \cite{shankar2010kaluza, shankar2012metric} for details.  

We start by visualizing the 5D space-time as  foliated by 4D hypersurfaces (with coordinates $x^{\mu}$) along the fifth dimension $x^{5}$, with metric 
\begin{equation}
ds_{5}^2 = \g_{\mu \nu}  dx^{\mu} dx^{\nu} + \g_{\mu 5} dx^{\mu} dx^{5} + \Phi^{2} [dx^5 ]^2 + \g_{\mu 5} \g_{\nu 5} \Phi^{-2} dx^{\mu} dx^{\nu} .
\label{5Dmetric}
\end{equation}
The metric is expressed in the above form for convenience because it turns out that once we impose the constraints in this theory, the 4D components $\g_{\mu \nu}$ play an identical role as the metric tensor in torsion-free GR at a kinematical level. Moreover, note that $\g_{\mu \nu}$ is exactly the induced metric on the hypersurfaces when $\g_{\mu 5}=0$.

The connection $\Christilde{}{}{} $ for this geometry is prescribed to ensure that any motion along the fifth dimension is unobservable and its 4D components are chosen to be torsionless as in GR. This is achieved by imposing the algebraic constraints $\Christilde{\mu}{\nu}{5} = \Christilde{\mu}{5}{5}=0$ and $\Christilde{\mu}{[ \alpha}{\beta ]}=0 $. 
By analyzing the geodesic equations, note that any motion along the fifth dimension does not affect the 4D components of the geodesic equations. These constraints along with the metricity condition completely determine all the components of torsion in terms of the metric.

Remarkably, this theory is indistinguishable from GR at a kinematic level. First, the 4D metric on the hypersurfaces, $\g_{\mu \nu}$, is independent of $x^{5}$, making all the hypersurfaces identical; this can be viewed as a consequence of imposing the fifth dimension to be hidden. Second, the components of the connection and the curvature tensor that are tangential to any hypersurface are identical to those in  GR with metric $\g_{\mu \nu}$. In effect, the terms contributed by the torsion exactly cancel off the terms contributed by the effect of the extra-dimension in evaluating the 4D components of connection and curvature.  Hence, hiding the extra dimension by requiring any motion along that dimension to not have any effect on the observable 4D motion, leads to a theory where the kinematic equations are identical to those in GR.  Moreover, this formalism does not require us to explicitly choose a topology for the extra dimension--it could be either a compact dimension or a large unbounded dimension. 
 
Since torsion is completely metric dependent in this framework, the field equations are derived by varying the action only with respect to the metric, and they take the form,\footnote{ In the original derivation of eq.~\ref{FE} presented in \cite{shankar2012metric}, the sign convention adopted to define the curvature and the Einstein tensor is different.}
\begin{eqnarray}
G^{\mu}_{\nu} -  \h^{\mu}_{\nu}=   \Sigma^{\mu}_{\nu} ,
\label{FE}
\\
\g^{\mu 5} [ \R_{\mu \nu} - \h_{\mu \nu}] = \Sigma^{5}_{\nu}  , \,\,  \frac{1}{2} \R=\Sigma^{5}_{5}.  \nonumber 
\end{eqnarray} 
Here $G^{\mu}_{\nu}$ is the standard Einstein tensor constructed from the 4D metric $\g_{\mu \nu}$ as in  GR, and $\R_{\mu \nu}$ and $\R$ are the torsion-free 4D Ricci tensor and Ricci scalar respectively constructed from the 4D metric $\g_{\mu \nu}$ as in  GR. The additional term $\h^{\mu}_{\nu}$ takes the form
\begin{equation}
\h^{\mu}_{\nu} =  \nabla_{\nu} \J^{\mu} - (\nabla . \J) \delta^{\mu}_{\nu} 
  + \J_{\nu} \J^{\mu}- (\J.\J)\delta^{\mu}_{\nu},
  \label{defH}
\end{equation}
where $\J_{\mu} \equiv  \Phi^{-1} \partial_{\mu} \Phi $, is  a 4 vector whose indices are raised and lowered with the 4D metric $\g_{\mu \nu}$ and its inverse. Similarly, the covariant derivative,  $\nabla_{\nu}$, is defined as in GR with Christoffel connection expressed in terms of the 4D metric $\g_{\mu \nu}$.
Note that the 4D components of the field equations only involve the 4D metric components $\g_{\mu \nu}$ and the scalar function $\Phi$ (which is simply $\g_{55}$), while the metric components $\g_{\mu 5}$ have completely decoupled out. 
Particularly, $\g_{\mu 5} =0$ and $\Sigma^{5}_{\nu}=0$ are always consistent solutions to the field equations.  Since only the 4D metric components $\g_{\mu \nu}$ and the 4D components of stress tensor $\Sigma^{\mu}_{\nu}$ are relevant for physical observations, eq.~\ref{FE} is sufficient to solve for all the physical degrees of freedom including $\Phi$. The unobservable extra dimensional stress tensor component $\Sigma^{5}_{5}$ can be defined to be $\mathcal{R}/2$ consistent with the field equations. 
Clearly, this theory reduces to GR when $\J_{\mu}$ vanishes. In GR, the Bianchi identity automatically implies the conservation of stress tensor, which is not true here because of the presence of the term $\h^{\mu}_{\nu}$ in eq.~\ref{FE}. Nevertheless, if we impose 4D matter conservation, $\nabla_{\mu}  \Sigma^{\mu}_{\nu} =0$, it would imply $\nabla_{\mu}  \h^{\mu}_{\nu} =0$ for self consistency of eq.~\ref{FE}.

\section{Static Spherical vacuum solutions}

To study the dynamics of a  spherically symmetric collapse in THED gravity, we need to solve the set of messy equations given in appendix-A derived from eq.~\ref{FE} with appropriate initial conditions. However the focus here is merely on the existence of compact non-singular solutions, particularly static spherically symmetric shells with a vacuum exterior. Let's express the static spherically symmetric 4D metric $\g_{\mu \nu}$  in the spherical polar form with  coordinates $(t,r,\theta, \phi)$ as 
\begin{equation} 
ds_{4}^2 =  \g_{\mu \nu}  dx^{\mu} dx^{\nu} = -A(r) dt^2 + B(r) dr^2 + r^2 [d\theta^{2} + \sin^2 \theta d\phi^2] .
\label{Sphericalmetric}
\end{equation}
In accordance with spherical symmetry, the metric component of the 5th dimension $\Phi$ (see eq.~\ref{5Dmetric}) should only depend on $r$,  and therefore $\J_r$ can be the only non vanishing component of $\J_{\mu}$ (see eq.~\ref{defH}).  

We start with the vacuum solutions ($\Sigma^{\mu}_{\nu}=0$) which are asymptotically flat with boundary conditions $A(r) \rightarrow 1$,  $B(r) \rightarrow 1$ and $\Phi(r) \rightarrow 1$ as $r \rightarrow \infty$. Taking $\g_{\mu 5}=0$, the 4D part of the field equations (eq.~\ref{FE}) can be solved to obtain 
\begin {equation}
 \frac{rA'(r)}{A(r)} = 2cf(r) ,  \,\,\, B(r) = - \frac{rf'(r)}{f(r)} = 1+2(1+c) f(r) + c f^2(r),
 \label{eq:sols}
\end{equation} 
where $f(r) \equiv r. \J_{r}$ and  $c$ is a free parameter that can range from $-\infty$ to $+\infty$. Here $(')$ corresponds to a derivative w.r.t. $r$. The function $\Phi(r)$ is expressible in terms of $A(r)$ by $A(r)= [\Phi(r) ]^{2c} $.  A detailed derivation of eq.~\ref{eq:sols} is given in \cite{shankar2012metric}.

Since $B(r) \rightarrow 1$ as $r \rightarrow \infty$, note that $f(r) \rightarrow 0$ such that $f(r) \propto 1/r$. The form of eq.~\ref{eq:sols} immediately reveals that the variable $r$ can be linearly scaled to leave the equations form-invariant. This scaling degree of freedom is similar to that in the Schwarzschild solution (with $A(r)=1/B(r)=1-2M/r $), where linearly scaling $r$ in the functions $A(r)$ and $B(r)$, simply amounts to linearly scaling the mass $M$. However, in this case the mass of the geometry depends not just on the scaling of $r$, but also on the parameter $c$.  By evaluating the gravitational acceleration at large $r$, we can arrive at a simple definition for the mass of this geometry  to be  
\begin{equation}
M =  c  \lim_{r \rightarrow \infty}  r f(r) .
\label{eq:Mdef}
\end{equation} 

In any static spherically symmetric geometry given by eq.~\ref{Sphericalmetric}, the radial acceleration of a static test particle at a radial coordinate $r$, as observed by a locally inertial observer is (see appendix-B)
\begin{equation}   
a(r) = \frac{A'(r)}{2A(r) \sqrt{B(r)}} .
\label{eq:acc}
\end{equation}
Clearly we require $A(r)$ and $B(r)$ to be positive, else the particle will be in a trapped region where it cannot be static. At large $r$, we expect this acceleration to be $M/r^2$ (Newtonian limit); thus the definition of mass in eq.~\ref{eq:Mdef} is a straightforward deduction from eqs.~\ref{eq:sols} and \ref{eq:acc}. 

The parameter $c$ does not have an explicit physical interpretation as it simply determines the gradient of the function $f(r)$ which is unobservable. The parameter $c$ determines how fast the scalar $\Phi$ asymptotes up or down to 1 for a given mass $M$. It is also worthy to note  that the Schwarzschild solution does not belong to the class of  solutions categorized by $c$, as the derivation of eq.~\ref{eq:sols} assumes that $f(r)$ is nonzero everywhere \cite{shankar2012metric}. 

\subsection*{Categorizing the solutions}

Other than the simple cases with $c=$ 0 or -1, we cannot obtain an analytic expression for $A(r)$, $B(r)$ and $f(r)$ from eq.~\ref{eq:sols}. To facilitate numerical computations of these functions, let's define $z\equiv 1/r$, so that the boundary conditions at $r=\infty$ is accessible at $z=0$. Further, with a redefinition $X \equiv 1/r f $, and treating $X$ as a function of $z$, \emph{i.e} $X(z)= z/f(1/z)$, we obtain from eq.~\ref{eq:sols}, 
\begin{equation}
-\frac{d X(z)}{d z}  = 2(1+c) + c \frac{z}{X(z)}  ,
\label{eq:X}
\end{equation}  
with the boundary condition that $X(z=0) = c/M$.
Once $X(z)$ is numerically computed, $f(r)$ is automatically obtained, and from eq.~\ref{eq:sols} we obtain $B(r)$ straightforwardly, and $A(r)$ is obtained to be 
\begin{equation}
A(r) = \exp \left[ -2 c \int_{0}^{1/r} \frac{dz}{X(z)} \right]  .
\end{equation} 

\subsubsection*{Special case $c=-1$ }

When $c=-1$, it turns out that 
\begin{eqnarray}
X(z)= - \frac{\sqrt{1+M^2 z^2}}{M} & \Longrightarrow & f(r)= \frac{- M/r}{\sqrt{1+M^2/r^2}}, \nonumber \\
A(r) & = &  \left( \sqrt{1+M^2/r^2} - M/r \right)^2 , \,\, B(r)= \frac{1}{1+M^2/r^2}  .
\end{eqnarray}

\subsubsection*{Special case $c=0$}

When $c=0$, we necessarily have $M=0$, and hence the boundary condition needed to solve eq.~\ref{eq:X} ($X(z=0) =c/M$) is meaningless. However, the equation can be solved by introducing an arbitrary constant $\alpha$, to obtain
\begin{equation}
X(z) = \alpha - 2 z  \Longrightarrow f(r)= \frac{1}{\alpha r -2}, \,\, A(r)= 1, \,\, B(r)= \frac{\alpha r}{\alpha r -2}  
\end{equation}  
Notice that for positive $\alpha$, $B(r)$ diverges at $r=2/\sqrt{\alpha}$, while it is well behaved for negative $\alpha$. Since our interest is in the positive mass ($M>0$) scenario, the case of $c=0$ is not of  further interest here.

\subsubsection*{Case $c>0$}

For any $c>0$, there is a point $z_o (=1/r_o)$  where $X(z_o) =0$. Consequently $f(r_o)=\infty$ and from eq.~\ref{eq:sols}, $B(r_o) =\infty$ and $A(r_o)=0$.  Noting that $X(z=0)=c/M$, the second term in the r.h.s. of eq.~\ref{eq:X} is negligible for $c \gg M$. Hence $X(z=0) - X(z_o) \simeq 2(1+c) (z_o - 0) $,  implying $r_o \simeq 2 M$ for $c \gg M$. The position of discontinuity ($r_o $) gradually increases as $c$ is decreased. 

The radial acceleration at $r_o$ given by  (see eq.~\ref{eq:acc}) is however finite with $a(r_o)=\sqrt{c}/r_o$; hence we might expect $r=r_o$ to be a coordinate singularity, and that the solution could be extended to the region $r<r_o$. However, it is impossible to self consistently extend the solution for $X(z)$ beyond $z=z_o$, from eq.~\ref{eq:X} without introducing an arbitrary discontinuity in $X(z)$ at $z=z_o$, which would then render the boundary conditions at infinity defining the mass $M$ meaningless. Moreover, since $A(r_o)=0$, we also have $\Phi(r_o)=0$, making the size of the extra dimension vanish at $r_o$. The fact that $r_o$ is larger than the Schwarzschild radius ($2M$) implies that it is not possible to construct a compact shell of a size smaller than what GR allows. Hence, the solutions with $c>0$ are not of any further interest to our examinations in this paper. 
 
 \subsubsection*{Case $c<0$}
 
When $c<0$, the functions $f(r)$, $A(r)$ and $B(r)$ are smooth for all $r>0$.  Since we require $M>0$, the function $f(r)$ is always negative and $f(r=0) = (\sqrt{1+c+c^2} - 1- c)/c $. The functions $A(r)$ and $B(r)$ are always positive and vanishing at $r=0$, that is $A(r=0)= B(r=0)=0$. However, the gravitational acceleration (eq.~\ref{eq:acc}) diverges at $r=0$, indicating a singularity. This can be verified by evaluating other scalar quantities derived from the curvature; for instance $R_{\mu \nu} R^{\mu \nu}$ diverges at $r=0$. Since $A(r)$ and $B(r)$ are always positive, there are no trapped surfaces in this geometry and there is no event horizon to censor the singularity. Thus the singularity at $r=0$ is a naked singularity \cite{shankar2012metric}. 

To understand why we obtain naked singularity solutions as opposed to black hole solutions, note that the vacuum  field equations (eq.~\ref{FE} with $\Sigma^{\mu}_{\nu} =0$) are exactly the GR field equations with $\h^{\mu}_{\nu}$ treated as an effective stress tensor induced from the extra dimensional interaction.  From eq.~\ref{defH} and eq.~\ref{eq:sols}, the components of $\h^{\mu}_{\nu}$ can be computed to be 
\begin{equation}
\h^{t}_{t} = \frac{cf^2(r)}{r^2 B(r)}, \, \h^{r}_{r} =  -\frac{f(r) (cf(r)+2) }{r^2 B(r)}, \,  
\h^{\theta}_{\theta} = \h^{\phi}_{\phi} =  \frac{f(r)}{r^2 B(r)} .
\end{equation}
This effective stress tensor violates the null energy condition (NEC), namely  $N \equiv \h^{\mu}_{\nu} U^{\nu} U_{\mu}$ is not always nonnegative for any null vector field $U^{\mu}$. The diagonal form of the metric (eq.~\ref{Sphericalmetric}) and the requirement that  $U^{\mu} U_{\mu} = 0$ leads to  
\begin{equation}
N= - \h^{t}_{t}  + \h^{r}_{r} x +\h^{\theta}_{\theta} y \,    \qquad \forall  \qquad  \{x,y \ge0, \,\,  x+y = 1 \}
\label{eq:N}
\end{equation}  
Notice that since $c$ and $f(r)$ are both negative, $\h^{t}_{t}$ and $\h^{\theta}_{\theta}$ are negative for all $r$, while $\h^{r}{_r}$ is positive for all $r$. To evaluate if $N$ becomes negative for any value of $x$ and $y$, we shall consider the extreme case with $x=0$ and $y = 1$ to obtain $N = (f(r) - c f^2(r))/r^2 B(r)$.  Since $f(r)$ is negative and falls-off to zero as $1/r$ for large $r$, note that $N$ will always become negative beyond some value of $r$. The exact value of $r$ beyond which $N$ becomes negative depends on $c$, and it should be computed numerically after solving eq.~\ref{eq:X}. It turns out that for all $-1<c<0$,  $N$ is negative for all $r$. While for larger negative values of $c$, the point beyond which $N$ becomes negative moves away from $r=0$ towards $r=3M$. Intriguingly, for $c< -1$, the effective stress tensor $\h^{\mu}_{\nu}$  does not violate NEC in the immediate neighborhood of r=0, and in some sense the solution with $c=-1$ is the limit where the effective stress tensor violates NEC everywhere.

However, since $\h^{\mu}_{\nu}$ is not the stress tensor of the observable matter, $N$ being negative does not bear any implication to the energy conditions satisfied by the matter. The absence of trapped surfaces gives rise to a possibility that with sufficiently large pressure, matter  can be compressed to arbitrarily small size, without forming a singularity while still satisfying NEC.  This is in stark contrast with GR where matter  cannot be compressed beyond its Schwarzschild radius without generating a singularity. In the reminder of the paper, we shall explore this possibility by constructing static shell solutions.

\section{Junction conditions for a static spherical shell} 

In this section we obtain simplified expressions for the energy density and pressure on a static spherically symmetric shell whose interior is flat and the exterior is given by the vacuum solutions described in the previous section. We first apply Israel's junction conditions \cite{israel1966singular}  on a static spherical shell in plain GR with external metric described by eq.~\ref{Sphericalmetric}, 
\begin{equation} 
ds_{4}^2 = -A(r) dt^2 + B(r) dr^2 + r^2 [d\theta^{2} + \sin^2 \theta d\phi^2], 
\end{equation}
and then derive the modifications introduced by the effect of extra dimension through the term $\h^{\mu}_{\nu}$ in the field equations (eq.~\ref{FE}). For clarity, we shall closely follow the procedure and notations used in  \cite{poisson2004relativist} in applying the junction conditions.

The first junction condition emphasizes that there should exist a local coordinate system in which the metric is continuous across the shell so as to yield a consistent induced metric on the shell from either sides. To be general, let $\{ x^{\alpha} \}$ denote the 4D coordinate system, and $\{ y^{a} \}$ denote the 3D coordinate system installed on the shell with induced metric $h_{a b}$. The $greek$ indices denote the 4D coordinates and the $latin$ indices denote the 3D coordinates. In our static spherically symmetric situation, clearly a coordinate system $y^{a}= \{ T(t),\theta,\phi \}$ can be installed on the shell from either sides of the shell. Let the shell exist at a radial coordinate $r=R$, with a surface density $\sigma$ and surface pressure $p$.

Now consider a  space-like geodesic congruence along the radially outward direction that orthogonally pierces the shell with $n_{\alpha}$ as its tangent vector at the shell.  Let $\ell$ be the affine parameter on these geodesics such that $\ell=0$ on the shell, $\ell<0$ in the interior and $\ell>0$ in the exterior. It is straightforward to see that the metric across the shell is continuous in the $x^{\alpha} = \{ T(t),\ell,\theta,\phi \}$ coordinate system. 
Since $n_{\alpha}$ is the unit normal to the shell surface, the extrinsic curvature of the shell is  given by 
\begin{equation}
K_{a b} = \nabla_{\beta}[ n_{\alpha}] \left[ \frac{\partial x^{\alpha}}{\partial y^{a}}\right] \left[ \frac{\partial x^{\beta}}{\partial y^{b}} \right]  .
\label{eq:Ecurv}
\end{equation}

Though the metric is continuous across the shell, its derivative is not continuous, and hence the extrinsic curvature will be discontinuous across the shell surface.  Moreover, since the Ricci tensor (and the Einstein tensor) is constructed from the second derivative of the metric, it would have a delta function singularity on the shell. If we write the metric  in the exterior as $g^{+}_{\mu \nu}$ and the metric in the interior as $g^{-}_{\mu \nu}$, then the overall metric can be written as 
\begin{equation}
g_{\mu \nu} = \Theta(-\ell) g^{-}_{\mu \nu}  + \Theta(+\ell)  g^{+}_{\mu \nu},
\end{equation}
where $\Theta(.)$ is the Heaviside function. Computing the Einstein tensor from this metric, 
\begin{equation}
G_{\mu \nu} =  \Theta(-\ell)  G^{-}_{\mu \nu} +  \Theta(+\ell)  G^{+}_{\mu \nu} 
 + \delta(\ell)   S_{\mu \nu}  . 
\label{eq:discont}
\end{equation}
Here  $G^{\pm}_{\mu \nu}$ is the Einstein tensor on either sides of the shell and $S_{\mu \nu}$  is the stress tensor on the shell. 
It turns out that  $S_{\mu \nu}$ is tangential to the shell, \emph{i.e.} $S_{\mu \nu} n^{\nu} =0$, and the non-vanishing components $S_{a b} = S_{\alpha \beta}  [ \partial x^{\alpha}/\partial y^{a}] [\partial x^{\beta}/\partial y^{b}]$ are given by 
\begin{equation}
S_{a b} = - [ (K^{+}_{a b} - K^{-}_{a b} ) - (K^{+} - K^{-})h_{ab} ]  .
\label{eq:S}
\end{equation}
Thus the surface stress tensor is given in terms of the difference in extrinsic curvature on either sides of the shell. The surface density and surface pressure on the shell are then $\sigma=-S^{T}_{T}$,  $p=S^{\theta}_{\theta}(= S^{\phi}_{\phi})$. 

We can straightforwardly apply this to the situation where the interior of the shell is flat, \emph{i.e.} $A(r)=B(r)=1$ for $r<R$, and the exterior is given by eq.~\ref{Sphericalmetric} for arbitrary positive functions $A(r)$ and $B(r)$ for $r>R$. In the $x^{\alpha}= \{ t, r, \theta, \phi \} $ coordinate system, note that $n_{\alpha} = \{0, \sqrt{B(R)},0,0 \} $, and with the 3D coordinate system on the shell $y^{a} =\{T(t),\theta,\phi \} $, note that  $T(t)= t \sqrt{A(R)}$.  Evaluating the extrinsic curvature on either sides of the shell from eq.~\ref{eq:Ecurv} and applying eq.~\ref{eq:S}, it turns out that 
\begin{equation}
\sigma = -S^{T}_{T} = \frac{2}{R}\left[1-  \frac{1}{\sqrt{B(R)}}  \right] ,  \,\,
p = S^{\theta}_{\theta}= \frac{1}{R}\left[ \frac{1+rA'(R)/2A(R)}{\sqrt{B(R)}} -1\right]   .
\label{EC}
\end{equation}

\subsection{Modified junction conditions } 

Let us now consider a thin shell of radius $R$ in THED gravity with a flat interior, and an exterior given by solutions described in the previous section. Based on the modified field equations (eq.~\ref{FE}), we can derive the modified junction conditions  to determine the surface density and surface pressure on the shell.  Note from eq.~\ref{defH} that $\h^{\mu}_{\nu}$ involves  derivatives of  $\J_{\mu}$. Since $\J_{\mu}$ vanishes in the interior but does not vanish in the exterior, it is discontinuous across the shell. Hence there would exist a delta function singularity in $\h^{\mu}_{\nu}$ just as the one in Einstein tensor in \ref{eq:discont}.  Since $\J^{\mu} = \Theta(-\ell) \J^{- \mu} + \Theta(+\ell) \J^{+ \mu}$, we can immediately see that $\partial_{\nu} \J^{\mu}$  will pick up a term $\delta(\ell) (\J^{+ \mu} - \J^{- \mu}) n_{\nu}$.  From the definition of $\h^{\mu}_{\nu}$ in eq.~\ref{defH} it is straightforward to pick out the delta function contribution to it. 
\begin{equation}
\h^{\,\,\mu}_{\nu} =  \Theta(-\ell) \h^{-\mu}_{\nu}+  \Theta(+\ell)  \h^{+\mu}_{\nu}
 + \delta(\ell)   [ (\J^{+ \mu} - \J^{- \mu} )n_{\nu} - (\J^{+\alpha} - \J^{- \alpha}) n_{\alpha} \delta^{\mu}_{\nu}]   .
\end{equation}

Since $\J_r$ and $n_r$ are the only  non vanishing components of $\J_{\mu}$ and $n_{\mu}$ respectively, it turns out that the delta function contribution in $\h^{T}_{T}$, $\h^{\theta}_{\theta}$ and $\h^{\phi}_{\phi}$ are all equal to 
\begin{equation}
 - (\J^{+ r} - \J^{- r} ) n_{r} = - \frac{f(R)}{R\sqrt{B(R)}},
\end{equation}
where $f(R)$ and $B(R)$ are the obtained from the vacuum exterior solution at the shell. Along with the delta function contribution from the Einstein tensor, this gives the total surface density and pressure.
\begin{equation}
\sigma =  \frac{2}{R}\left[1-\frac{1}{\sqrt{B(R)}} \right]  -\frac{f(R)}{R\sqrt{B(R)}},  \,\,
p = \frac{1}{R}\left[ \frac{1+rA'(R)/2A(R)}{\sqrt{B(R)}} -1\right] + \frac{f(R)}{R\sqrt{B(R)}}  .
\label{MEC}
\end{equation}

\section{Evaluating NEC on the Shell}

Now we are equipped to evaluate if the NEC is valid on arbitrarily  small shells producing a massive exterior geometry.   For a given mass $M$ of the exterior geometry, here we numerically evaluate the exterior metric and compute the surface density $\sigma$  and pressure $p$ of the shell. The NEC requires that $\sigma + p \ge 0$. 

Since the vacuum solutions in THED with $c<0$ do not have trapped surfaces close to the singularity, the shell can be placed arbitrarily close to $r=0$. As a comparison, we shall also consider such possibilities within the context of GR. In the Schwarzschild solution, the singularity is surrounded by trapped surfaces. But in an electrically charged black hole given by Reisner-Nordstorm (RN) metric, there are no trapped surfaces in the immediate neighborhood of the singularity. Similarly, with scalar charge, the solution given by JNW metric exhibits a naked singularity with no trapped surfaces. Hence it is in principle possible to consider static shells arbitrarily close to $r=0$ with RN or JNW exterior geometries. However it is worth noting that the exterior geometries of RN and JNW  shells are not vacuum, but they will contain electric and scalar fields respectively.  

For the RN geometry, the metric components are  $A(r) = 1/ B(r) = (1- 2M/r +Q^2/r^2) $. There are trapped surfaces between the external horizon and the internal horizon located at $r= M+\sqrt{M^2-Q^2}$ and $r=M-\sqrt{M^2-Q^2}$ respectively. A singularity-free static shell constructed within the internal horizon must necessarily violate NEC because the exterior geometry will have trapped surfaces, else there will be a contradiction to the singularity theorem \cite{penrose1965gravitational}.

\begin{figure} 
\begin{center}
\includegraphics[width=1 \textwidth]{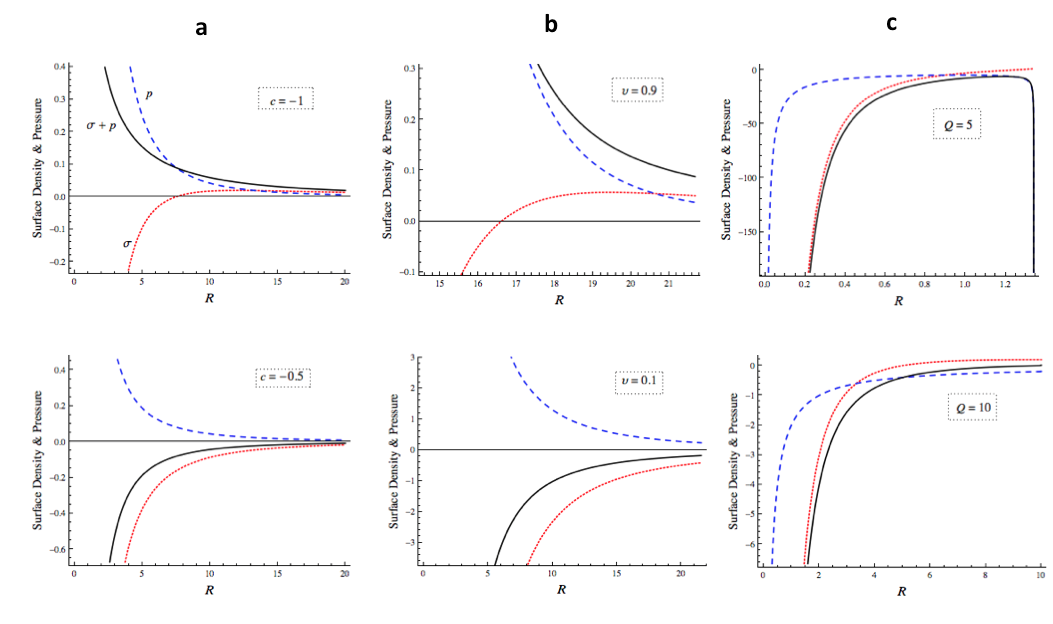}
\caption{Surface density and pressure of  compact static shells are plotted as function of the shell's radius. The exterior geometry outside the shell is given by THED solutions in \textbf{a}, JNW solutions in \textbf{b} and RN solutions in \textbf{c}. Dashed curves correspond to the pressure, and the dotted curves correspond to the density. The solid curves show the sum of density and pressure, whose positivity implies that NEC is satisfied. 
}
\label{NECfig}
\end{center}
\end{figure}

The JNW metric \cite{janis1968reality} is written as
\begin{equation}
ds_{jnw}^2 = -(1-b/ \x)^{\nu} dt^2 + (1-b/ \x)^{-\nu} d \x^2 + (1-b/ \x)^{1-\nu} \x^2 [d\theta^{2} + \sin^2 \theta d\phi^2], 
\end{equation}
where $\nu =2M/b$ and $b=2\sqrt{M^2+q^2}$, and $q$ is the scalar charge. The parameter $\nu$ can take values between 0 and 1. With $q=0$, we have $\nu=1$, giving back the Schwarzschild solution. For any $\nu<1$, there is a naked singularity at $\x= b$. However, when we convert to the geometric radial coordinate $r$ from $\x$, we have 
\begin{equation}
\x^2 (1-b/ \x)^{1-\nu} = r^2 ,
\end{equation}
and the singularity is indeed at the center $r=0$. Implicitly evaluating $\x$ as a function of $r$ from the above equation, we obtain the metric components to be
\begin{equation}
A(r)=  (1-b/ \x (r) )^{\nu} , \qquad  B(r)=  (1-b/ \x (r))^{-\nu} \left[ \frac{ d \x (r)}{d r} \right]^2 ,
\end{equation}
which are always positive. Since there are no trapped surfaces, arbitrarily small shells can be considered with JNW exterior, and we can evaluate the NEC on the shells by computing $\sigma$ and $p$ from eq.~\ref{EC}.

In figure.~\ref{NECfig},  we numerically evaluate and plot $\sigma$, $p$, and $\sigma+p$ as a function of the radius of the shell $R$. For comparison, in all cases the mass has been fixed to $M=10$, and our focus is on compact shells with $R$ smaller than the Schwarzschild radius ($R<2M$); for shells with RN exterior we focus on $R$ smaller than the inner horizon. Fig.~\ref{NECfig}a shows the plots with exterior metric given by THED solutions for $c=-0.5$ and $c=-1$; fig.~\ref{NECfig}b shows the plots with exterior JNW metric for $\nu=0.9$ and $\nu=0.1$; fig.~\ref{NECfig}c shows the plots with exterior RN metric for $Q=5$ and  $Q=10$ (the extremal black hole).   

(i) For exterior geometry given by THED solutions, NEC is always satisfied for $c \le -1$ irrespective of how small the shell is. But for $0 > c > -1$, the NEC is violated for small shells, as exemplified by the solid curve representing $\sigma+p$ in the lower panel of fig.~\ref{NECfig}a.  Although the NEC is always satisfied for $c \le -1$, the density $\sigma$ is negative for small shells, implying that weak energy condition is violated. 
The case of $c=-1$  appears to be special because : no matter how small the shell is, it satisfies the NEC, while at the same time the extradimensionally induced stress tensor $\h^{\mu}_{\nu}$ violates the NEC everywhere (see eq.~\ref{eq:N}). 

(ii) For exterior geometry given by JNW metric, when the charge $q$ is large ($\nu$ is close to zero), NEC is violated for small shells, while NEC is satisfied for all shells when $\nu$ is larger.  

(iii) For RN exterior geometry, NEC is always violated for shells inside the inner horizon, as expected from the singularity theorem. For shells outside the outer horizon (not shown in fig.~\ref{NECfig}c),  both density and pressure are positive, satisfying the NEC.   

\section{Discussion}

In classical theories of gravity where massive vacuum solutions are naked singularities rather than black holes with event horizons that shield the singularity, gravitational collapse can  in principle stop at a point where the pressure is very large and the matter is compressed to a super-compact state. The static shell solutions constructed here demonstrates this possibility. It is however important to not misconstrue fig.~\ref{NECfig} as depicting how the surface density and pressure of a slowly collapsing shell would evolve as the size of the shell shrinks. The solutions constructed here are purely static and they just point to the possibility that a collapse could result in such compact non-singular solutions.
To confirm that a spherically symmetric collapse would lead to  a compact non-singular end state, the dynamical equations given in appendix-A has to be solved. The initial conditions of collapse would play a significant role in determining the value of $c$ that describes the vacuum exterior geometry (given by eq.~\ref{eq:sols}). Further investigations are needed in a dynamical set up to ensure that initial conditions with  a fixed value of $c$ ($\le -1$) for the exterior geometry would remain that way as the collapse evolves.

Classical matter is generally expected to have positive energy density and pressure, consistent with our observations at moderate energy scales.  Although classical scalar fields with non-minimal coupling to gravity can violate NEC under tailored conditions \cite{visser1999energy}, no observed matter has  really shown this violation. It is well understood that quantum effects can lead to negative energy density (thus violating the weak energy condition), but it is not expected to violate the NEC; \emph{i.e.}, even if $\sigma$ is negative, $\sigma +p$ should be non-negative. The solutions to $\sigma+p$ plotted in fig.~\ref{NECfig}a,b (top panel) are examples illustrating NEC satisfying super compact matter distribution. By conventional definition,  \emph{non-exotic} matter essentially satisfies the NEC. Here we have simply demonstrated that in THED gravity, gravitational collapse singularities can potentially be avoided without availing any exotic matter.

Since THED gravity matches GR at a kinematic level, all the equations of motion of test particles will be identical to that in GR for a given 4D metric. So, in the weak field regime  (any test within the solar system), it is unlikely to observationally distinguish THED gravity from GR. However in the strong field regime, the existence of the modification term $\h^{\mu}_{\nu}$ in the field equations (eq.~\ref{FE}) can play a significant role in the solution to the 4D metric. The existence of naked singularities or super-compact objects without trapped surfaces can lead to a very distinct signature in gravitational lensing phenomena.   Such signatures have been calculated  in the context of JNW naked singularities \cite{virbhadra2002gravitational}. Similarly non-singular super compact objects in THED, that may appear as  black holes from large distances, might show a distinct fingerprint in gravitational waves if two such objects were to merge.   
It is also noteworthy that cosmological singularity (big bang) can be avoided in THED gravity because it yields  oscillatory solutions for the universe \cite{shankar2012metric} (the term $\h^{\mu}_{\nu}$ dominates the field equations when the universe is very small). With detailed perturbation analysis of the cosmological solutions, we could potentially find fingerprints of THED in the cosmic microwave background. 

I conclude by noting that  modified semiclassical theories of gravity (like THED) have the potential to eradicate  gravitational singularities without having to quantize gravity.

\textbf{Acknowledgements}: The author thanks funding from National Science Foundation grant, NSF BCS- 1058937, and support from the Aspen Center for Physics and their NSF Grant- 1066293.

\section*{References}


\begin{thebibliography}{22}%
\makeatletter
\providecommand \@ifxundefined [1]{%
 \@ifx{#1\undefined}
}%
\providecommand \@ifnum [1]{%
 \ifnum #1\expandafter \@firstoftwo
 \else \expandafter \@secondoftwo
 \fi
}%
\providecommand \@ifx [1]{%
 \ifx #1\expandafter \@firstoftwo
 \else \expandafter \@secondoftwo
 \fi
}%
\providecommand \natexlab [1]{#1}%
\providecommand \enquote  [1]{``#1''}%
\providecommand \bibnamefont  [1]{#1}%
\providecommand \bibfnamefont [1]{#1}%
\providecommand \citenamefont [1]{#1}%
\providecommand \href@noop [0]{\@secondoftwo}%
\providecommand \href [0]{\begingroup \@sanitize@url \@href}%
\providecommand \@href[1]{\@@startlink{#1}\@@href}%
\providecommand \@@href[1]{\endgroup#1\@@endlink}%
\providecommand \@sanitize@url [0]{\catcode `\\12\catcode `\$12\catcode
  `\&12\catcode `\#12\catcode `\^12\catcode `\_12\catcode `\%12\relax}%
\providecommand \@@startlink[1]{}%
\providecommand \@@endlink[0]{}%
\providecommand \url  [0]{\begingroup\@sanitize@url \@url }%
\providecommand \@url [1]{\endgroup\@href {#1}{\urlprefix }}%
\providecommand \urlprefix  [0]{URL }%
\providecommand \Eprint [0]{\href }%
\providecommand \doibase [0]{http://dx.doi.org/}%
\providecommand \selectlanguage [0]{\@gobble}%
\providecommand \bibinfo  [0]{\@secondoftwo}%
\providecommand \bibfield  [0]{\@secondoftwo}%
\providecommand \translation [1]{[#1]}%
\providecommand \BibitemOpen [0]{}%
\providecommand \bibitemStop [0]{}%
\providecommand \bibitemNoStop [0]{.\EOS\space}%
\providecommand \EOS [0]{\spacefactor3000\relax}%
\providecommand \BibitemShut  [1]{\csname bibitem#1\endcsname}%
\let\auto@bib@innerbib\@empty
\bibitem [{\citenamefont {Penrose}(1965)}]{penrose1965gravitational}%
  \BibitemOpen
  \bibfield  {author} {\bibinfo {author} {\bibfnamefont {R.}~\bibnamefont
  {Penrose}},\ }\href@noop {} {\bibfield  {journal} {\bibinfo  {journal}
  {Physical Review Letters}\ }\textbf {\bibinfo {volume} {14}},\ \bibinfo
  {pages} {57} (\bibinfo {year} {1965})}\BibitemShut {NoStop}%
\bibitem [{\citenamefont {Senovilla}\ and\ \citenamefont
  {Garfinkle}(2014)}]{senovilla20141965}%
  \BibitemOpen
  \bibfield  {author} {\bibinfo {author} {\bibfnamefont {J.~M.}\ \bibnamefont
  {Senovilla}}\ and\ \bibinfo {author} {\bibfnamefont {D.}~\bibnamefont
  {Garfinkle}},\ }\href@noop {} {\bibfield  {journal} {\bibinfo  {journal}
  {arXiv preprint arXiv:1410.5226}\ } (\bibinfo {year} {2014})}\BibitemShut
  {NoStop}%
\bibitem [{\citenamefont {Penrose}(1999)}]{penrose1999question}%
  \BibitemOpen
  \bibfield  {author} {\bibinfo {author} {\bibfnamefont {R.}~\bibnamefont
  {Penrose}},\ }\href@noop {} {\bibfield  {journal} {\bibinfo  {journal}
  {Journal of Astrophysics and Astronomy}\ }\textbf {\bibinfo {volume} {20}},\
  \bibinfo {pages} {233} (\bibinfo {year} {1999})}\BibitemShut {NoStop}%
\bibitem [{\citenamefont {Shapiro}\ and\ \citenamefont
  {Teukolsky}(1991)}]{shapiro1991formation}%
  \BibitemOpen
  \bibfield  {author} {\bibinfo {author} {\bibfnamefont {S.~L.}\ \bibnamefont
  {Shapiro}}\ and\ \bibinfo {author} {\bibfnamefont {S.~A.}\ \bibnamefont
  {Teukolsky}},\ }\href@noop {} {\bibfield  {journal} {\bibinfo  {journal}
  {Physical review letters}\ }\textbf {\bibinfo {volume} {66}},\ \bibinfo
  {pages} {994} (\bibinfo {year} {1991})}\BibitemShut {NoStop}%
\bibitem [{\citenamefont {Wald}(1999)}]{wald1999gravitational}%
  \BibitemOpen
  \bibfield  {author} {\bibinfo {author} {\bibfnamefont {R.~M.}\ \bibnamefont
  {Wald}},\ }in\ \href@noop {} {\emph {\bibinfo {booktitle} {Black Holes,
  Gravitational Radiation and the Universe}}}\ (\bibinfo  {publisher}
  {Springer},\ \bibinfo {year} {1999})\ pp.\ \bibinfo {pages}
  {69--86}\BibitemShut {NoStop}%
\bibitem [{\citenamefont {Janis}\ \emph {et~al.}(1968)\citenamefont {Janis},
  \citenamefont {Newman},\ and\ \citenamefont {Winicour}}]{janis1968reality}%
  \BibitemOpen
  \bibfield  {author} {\bibinfo {author} {\bibfnamefont {A.~I.}\ \bibnamefont
  {Janis}}, \bibinfo {author} {\bibfnamefont {E.~T.}\ \bibnamefont {Newman}}, \
  and\ \bibinfo {author} {\bibfnamefont {J.}~\bibnamefont {Winicour}},\
  }\href@noop {} {\bibfield  {journal} {\bibinfo  {journal} {Physical Review
  Letters}\ }\textbf {\bibinfo {volume} {20}},\ \bibinfo {pages} {878}
  (\bibinfo {year} {1968})}\BibitemShut {NoStop}%
\bibitem [{\citenamefont {Virbhadra}\ and\ \citenamefont
  {Ellis}(2002)}]{virbhadra2002gravitational}%
  \BibitemOpen
  \bibfield  {author} {\bibinfo {author} {\bibfnamefont {K.}~\bibnamefont
  {Virbhadra}}\ and\ \bibinfo {author} {\bibfnamefont {G.~F.}\ \bibnamefont
  {Ellis}},\ }\href@noop {} {\bibfield  {journal} {\bibinfo  {journal}
  {Physical Review D}\ }\textbf {\bibinfo {volume} {65}},\ \bibinfo {pages}
  {103004} (\bibinfo {year} {2002})}\BibitemShut {NoStop}%
\bibitem [{\citenamefont {Shankar}\ \emph {et~al.}(2012)\citenamefont
  {Shankar}, \citenamefont {Balaraman},\ and\ \citenamefont
  {Wali}}]{shankar2012metric}%
  \BibitemOpen
  \bibfield  {author} {\bibinfo {author} {\bibfnamefont {K.~H.}\ \bibnamefont
  {Shankar}}, \bibinfo {author} {\bibfnamefont {A.}~\bibnamefont {Balaraman}},
  \ and\ \bibinfo {author} {\bibfnamefont {K.~C.}\ \bibnamefont {Wali}},\
  }\href@noop {} {\bibfield  {journal} {\bibinfo  {journal} {Physical Review
  D}\ }\textbf {\bibinfo {volume} {86}},\ \bibinfo {pages} {024007} (\bibinfo
  {year} {2012})}\BibitemShut {NoStop}%
\bibitem [{\citenamefont {Curiel}(2014)}]{curiel2014primer}%
  \BibitemOpen
  \bibfield  {author} {\bibinfo {author} {\bibfnamefont {E.}~\bibnamefont
  {Curiel}},\ }\href@noop {} {\bibfield  {journal} {\bibinfo  {journal} {arXiv
  preprint arXiv:1405.0403}\ } (\bibinfo {year} {2014})}\BibitemShut {NoStop}%
\bibitem [{\citenamefont {Mathur}(2005)}]{mathur2005fuzzball}%
  \BibitemOpen
  \bibfield  {author} {\bibinfo {author} {\bibfnamefont {S.~D.}\ \bibnamefont
  {Mathur}},\ }\href@noop {} {\bibfield  {journal} {\bibinfo  {journal}
  {Fortschritte der Physik}\ }\textbf {\bibinfo {volume} {53}},\ \bibinfo
  {pages} {793} (\bibinfo {year} {2005})}\BibitemShut {NoStop}%
\bibitem [{\citenamefont {Balasubramanian}\ \emph {et~al.}(2008)\citenamefont
  {Balasubramanian}, \citenamefont {Gimon},\ and\ \citenamefont
  {Levi}}]{balasubramanian2008four}%
  \BibitemOpen
  \bibfield  {author} {\bibinfo {author} {\bibfnamefont {V.}~\bibnamefont
  {Balasubramanian}}, \bibinfo {author} {\bibfnamefont {E.~G.}\ \bibnamefont
  {Gimon}}, \ and\ \bibinfo {author} {\bibfnamefont {T.~S.}\ \bibnamefont
  {Levi}},\ }\href@noop {} {\bibfield  {journal} {\bibinfo  {journal} {Journal
  of High Energy Physics}\ }\textbf {\bibinfo {volume} {2008}},\ \bibinfo
  {pages} {056} (\bibinfo {year} {2008})}\BibitemShut {NoStop}%
\bibitem [{\citenamefont {Rama}(2013)}]{rama2013m}%
  \BibitemOpen
  \bibfield  {author} {\bibinfo {author} {\bibfnamefont {S.~K.}\ \bibnamefont
  {Rama}},\ }\href@noop {} {\bibfield  {journal} {\bibinfo  {journal} {Physical
  Review D}\ }\textbf {\bibinfo {volume} {88}},\ \bibinfo {pages} {044007}
  (\bibinfo {year} {2013})}\BibitemShut {NoStop}%
\bibitem [{\citenamefont {Rama}(2011)}]{rama2011static}%
  \BibitemOpen
  \bibfield  {author} {\bibinfo {author} {\bibfnamefont {S.~K.}\ \bibnamefont
  {Rama}},\ }\href@noop {} {\bibfield  {journal} {\bibinfo  {journal} {arXiv
  preprint arXiv:1111.1897}\ } (\bibinfo {year} {2011})}\BibitemShut {NoStop}%
\bibitem [{\citenamefont {Rama}(2014)}]{rama2014massive}%
  \BibitemOpen
  \bibfield  {author} {\bibinfo {author} {\bibfnamefont {S.~K.}\ \bibnamefont
  {Rama}},\ }\href@noop {} {\bibfield  {journal} {\bibinfo  {journal} {arXiv
  preprint arXiv:1409.3462}\ } (\bibinfo {year} {2014})}\BibitemShut {NoStop}%
\bibitem [{\citenamefont {Israel}(1966)}]{israel1966singular}%
  \BibitemOpen
  \bibfield  {author} {\bibinfo {author} {\bibfnamefont {W.}~\bibnamefont
  {Israel}},\ }\href@noop {} {\bibfield  {journal} {\bibinfo  {journal} {Il
  Nuovo Cimento B Series 10}\ }\textbf {\bibinfo {volume} {44}},\ \bibinfo
  {pages} {1} (\bibinfo {year} {1966})}\BibitemShut {NoStop}%
\bibitem [{\citenamefont {Hehl}\ \emph {et~al.}(1976)\citenamefont {Hehl},
  \citenamefont {Von~der Heyde}, \citenamefont {Kerlick},\ and\ \citenamefont
  {Nester}}]{hehl1976general}%
  \BibitemOpen
  \bibfield  {author} {\bibinfo {author} {\bibfnamefont {F.~W.}\ \bibnamefont
  {Hehl}}, \bibinfo {author} {\bibfnamefont {P.}~\bibnamefont {Von~der Heyde}},
  \bibinfo {author} {\bibfnamefont {G.~D.}\ \bibnamefont {Kerlick}}, \ and\
  \bibinfo {author} {\bibfnamefont {J.~M.}\ \bibnamefont {Nester}},\
  }\href@noop {} {\bibfield  {journal} {\bibinfo  {journal} {Reviews of Modern
  Physics}\ }\textbf {\bibinfo {volume} {48}},\ \bibinfo {pages} {393}
  (\bibinfo {year} {1976})}\BibitemShut {NoStop}%
\bibitem [{\citenamefont {Olmo}\ and\ \citenamefont
  {Rubiera-Garcia}(2013)}]{olmo2013importance}%
  \BibitemOpen
  \bibfield  {author} {\bibinfo {author} {\bibfnamefont {G.~J.}\ \bibnamefont
  {Olmo}}\ and\ \bibinfo {author} {\bibfnamefont {D.}~\bibnamefont
  {Rubiera-Garcia}},\ }\href@noop {} {\bibfield  {journal} {\bibinfo  {journal}
  {Physical Review D}\ }\textbf {\bibinfo {volume} {88}},\ \bibinfo {pages}
  {084030} (\bibinfo {year} {2013})}\BibitemShut {NoStop}%
\bibitem [{\citenamefont {Pop{\l}awski}(2012)}]{poplawski2012nonsingular}%
  \BibitemOpen
  \bibfield  {author} {\bibinfo {author} {\bibfnamefont {N.}~\bibnamefont
  {Pop{\l}awski}},\ }\href@noop {} {\bibfield  {journal} {\bibinfo  {journal}
  {Physical Review D}\ }\textbf {\bibinfo {volume} {85}},\ \bibinfo {pages}
  {107502} (\bibinfo {year} {2012})}\BibitemShut {NoStop}%
\bibitem [{\citenamefont {Poplawski}(2014)}]{poplawski2014universe}%
  \BibitemOpen
  \bibfield  {author} {\bibinfo {author} {\bibfnamefont {N.~J.}\ \bibnamefont
  {Poplawski}},\ }\href@noop {} {\bibfield  {journal} {\bibinfo  {journal}
  {arXiv preprint arXiv:1410.3881}\ } (\bibinfo {year} {2014})}\BibitemShut
  {NoStop}%
\bibitem [{\citenamefont {Shankar}\ and\ \citenamefont
  {Wali}(2010)}]{shankar2010kaluza}%
  \BibitemOpen
  \bibfield  {author} {\bibinfo {author} {\bibfnamefont {K.~H.}\ \bibnamefont
  {Shankar}}\ and\ \bibinfo {author} {\bibfnamefont {K.~C.}\ \bibnamefont
  {Wali}},\ }\href@noop {} {\bibfield  {journal} {\bibinfo  {journal} {Modern
  Physics Letters A}\ }\textbf {\bibinfo {volume} {25}},\ \bibinfo {pages}
  {2121} (\bibinfo {year} {2010})}\BibitemShut {NoStop}%
\bibitem [{\citenamefont {Poisson}(2004)}]{poisson2004relativist}%
  \BibitemOpen
  \bibfield  {author} {\bibinfo {author} {\bibfnamefont {E.}~\bibnamefont
  {Poisson}},\ }\href@noop {} {\emph {\bibinfo {title} {A relativist's toolkit:
  the mathematics of black-hole mechanics}}}\ (\bibinfo  {publisher} {Cambridge
  University Press},\ \bibinfo {year} {2004})\BibitemShut {NoStop}%
\bibitem [{\citenamefont {Visser}\ and\ \citenamefont
  {Barcelo}(1999)}]{visser1999energy}%
  \BibitemOpen
  \bibfield  {author} {\bibinfo {author} {\bibfnamefont {M.}~\bibnamefont
  {Visser}}\ and\ \bibinfo {author} {\bibfnamefont {C.}~\bibnamefont
  {Barcelo}},\ }\href@noop {} {\bibfield  {journal} {\bibinfo  {journal} {arXiv
  preprint gr-qc/0001099}\ } (\bibinfo {year} {1999})}\BibitemShut {NoStop}%
\end{thebibliography}%


%

\section*{Appendix}

\subsection*{A: Time-dependent Spherically Symmetric Equations}

We need to study the dynamics of collapse to properly evaluate whether THED gravity will generally avoid collapse singularities. To study the time dependent spherically symmetric equations in THED gravity, consider the extra dimensional metric field $\Phi(r,t)$ and the time dependent 4D metric given by  
\begin{equation} 
ds_{4}^2 = -A(r,t) dt^2 + B(r,t) dr^2 + r^2 [d\theta^{2} + \sin^2 \theta d\phi^2] 
\end{equation} 
Using an over-dot to represent time derivative and an apostrophe to represent derivative with $r$, and suppressing the functional dependence of the terms w.r.t to $r$ and $t$, the field equations (eq.~\ref{FE}) take the form

\begin{equation}
-2 A B r^2 \Phi''  -  4 r A B \Phi'  + A B' \Phi' r^2  + 2 \Phi A B' r + \dot{\Phi} \dot{B} r^2 B + 2 A B \Phi (B-1)  = (- \Sigma_{t}^{t} ) 2 A \Phi B^2 r^2   
\label{E1}
\end{equation}

\begin{equation}
2 r A B \dot{\Phi}'  -  2 \Phi A \dot{B}  - r \Phi' A \dot{B}  - r \dot{\Phi} A' B  =   (- \Sigma_{t}^{r} )  2 A \Phi B^2 r 
\label{E2}
\end{equation}

\begin{equation}
2 A B r^2 \ddot{\Phi} - r^2 A A'  \Phi'  -4 r A^2 \Phi' - 2 A A' r \Phi - r^2 \dot{A}B \dot{\Phi} + 2 A^2 \Phi (B-1)  =   (- \Sigma_{r}^{r} ) 2 A^2  B  \Phi r^2 
\label{E3}
\end{equation} 

\begin{eqnarray}
&& -2 A B \Phi r^2 A'' + 2 A B \Phi r^2 \ddot{B}  + \Phi B r^2 A'^{2} + A \Phi r^2 A' B' + 2 A B \Phi r A' -\Phi r^2 A \dot{B}^2 
\nonumber \\
&&  - \Phi r^2 B \dot{A} \dot{B} - 2 A^2 \Phi r B' + 12 A^2 B r \Phi' - 8 A^2 B \Phi (B-1)  =  (- \Sigma_{\theta}^{\theta} ) 4 A^2 B^2  \Phi r^2
\label{E4}
\end{eqnarray}

If one were to consider a perfect fluid solution, then the density $\rho(r,t) = -\Sigma_{t}^{t} (r,t)$ and the pressure $p(r,t) = \Sigma_{r}^{r} (r,t)=  \Sigma_{\theta}^{\theta} (r,t)=  \Sigma_{\phi}^{\phi} (r,t)$, and $\Sigma_{t}^{r}(r,t) =0$.  It is useful to know if the equivalent of Birkhoff's theorem is valid--when the matter has collapsed into a spherical region outside which the stress tensor is identically zero, is the external geometry static? Although the complexity of the above equations makes it extremely difficult to prove such a theorem, it appears reasonable to expect the external geometry to be static since conservation of matter heuristically implies that the  mass of the geometry is fixed.

\subsection*{B: Acceleration of a static test particle}

Let a test particle be held fixed in a static spherically symmetric geometry
\begin{equation}
ds_{4}^{2} = - A(r) dt^{2} + B(r) dr^{2} + r^{2} [d\theta^{2} + \sin^2 \theta d\phi^2] , 
\end{equation}
where  $A(r)$ and $B(r)$ are assumed to be positive. An external force is required to hold this particle at rest, implying it has to accelerate in a locally inertial frame to stay static. 
To compute this acceleration, consider a locally inertial observer momentarily at rest at the position of the particle. Let the four velocity of the observer's geodesic be  $U^{\mu}(\tau)$, where $\tau$ is the proper time along the geodesic. By considering space-like geodesics orthogonal to it, a Fermi Normal (FN) coordinate system can be erected such that the particle is at rest at the origin of the FN system. Due to spherical symmetry, it is sufficient to consider just radial geodesics to compute the acceleration of the test particle. Let $U^{\mu}(\tau)$ be a radial time-like geodesic, so the only non-vanishing components are $U^{r}(\tau)$ and $U^{t}(\tau)$, and $U^{r}(0)=0$. For each $\tau$, there exists a space-like radial geodesic $V^{\mu}(s)$ which is orthogonal to $U^{\mu}(\tau)$ at $s=0$. Here $s$ is the affine parameter on the space-like geodesic. The FN coordinate system is thus parametrized by $(\tau,s)$.

For convenience let us denote the observer's radial coordinate at any $\tau$ by $R_o (\tau)$, so that the test particle is located at $R_o(0)$. At $\tau , s = 0$, the orthonormality condition implies   
\begin{equation}
- (U^{t})^2 A + (U^{r})^2  B =-1,\,\,\,\,
- (V^{t})^2 A +  (V^{r})^2 B =+1,\,\,\,\,
- (U^{t}V^{t}) A + (U^{r} V^{r}) B =0.
 \end{equation}
 Here  $A$ and $B$ are understood to be the functions $A(r)$ and $B(r)$ evaluated at $r=R_o(0)$. Since $U^{r}=0$ at $\tau=0$, the above conditions imply $U^{t} =1/\sqrt{A}$ and $V^{r} =1/\sqrt{B}$. For the purposes of computing the momentary acceleration of the particle in the FN coordinates, it is sufficient  to consider infinitesimally small values of $\tau$ and $s$. Using the geodesic equation, we can obtain $U^{r}(\tau)$ for small $\tau$.
\begin{equation}
\frac{d}{d\tau} U^{t} +(U^{t})^2  \Chris{t}{t}{t}  =0,  \qquad   \frac{d}{d\tau} U^{r} + (U^{t})^2 \Chris{r}{t}{t}  =0  .
\end{equation}
The affine connection $\Chris{}{}{}$ is assumed to be evaluated at $r=R_o(0)$. Since $U^{r} (\tau)= d R_o(\tau)/d\tau$ and $U^{r}(0)=0$, we have
\begin{equation}
U^{r} (\tau)= - A^{-1} \Chris{r}{t}{t} \tau \,\, \Rightarrow \,\, 
R_o(\tau) = R_o(0) -\frac{1}{2} A^{-1}\Chris{r}{t}{t}  \tau^2 .
\end{equation}
Now, at any $\tau$, if we denote the radial coordinate of the space-like geodesic as $R_p(s)$,  then $V^{r} (s)= d R_p(s)/ds$. Since $\tau$ is very small, $V^{r}(s=0) =1/\sqrt{B}$ up to leading order. Hence for very small values of $s$ and $\tau$,
\begin{equation}
R_p(s) = R_o(\tau) +  s/ \sqrt{B}  .  
\end{equation}
 
The position of the test particle in the global coordinate system is fixed at $r=R_o(0)$. The particle's position in the  FN coordinate system can be obtained by simply setting $R_p(s)=R_o(0)$, which yields  
\begin{equation}
s = \frac{1}{2} \left[ \frac{ \sqrt{B}}{A} \Chris{r}{t}{t} \right]  \tau^2 .
\end{equation}
The acceleration of the test particle in the FN coordinates is then $a= \Chris{r}{t}{t} (\sqrt{B}/A)$.
In the spherically symmetric metric $\Chris{r}{t}{t}=A'/2B$, hence
\begin{equation}
a= \frac{A'}{2A\sqrt{B}}  .
\end{equation}

\end{document}